\documentclass[11pt,a4paper]{article}

\pdfoutput=1
\usepackage{jheppub}
\usepackage{graphicx,color,float}
\usepackage{hyperref}
\usepackage{multirow}
\usepackage{verbatim}
\usepackage{longtable}
\usepackage{amsmath}
\usepackage{amsfonts}
\usepackage{amssymb}
\usepackage{rotating}

\usepackage[normalem]{ulem}

\usepackage{siunitx}

\usepackage{orcidlink}

\def\gsim{\raise0.3ex\hbox{$\;>$\kern-0.75em\raise-1.1ex\hbox{$\sim\;$}}}
\def\lsim{\raise0.3ex\hbox{$\;<$\kern-0.75em\raise-1.1ex\hbox{$\sim\;$}}}

\begin{document}


\title{Searching for heavy neutral leptons coupled to axion-like particles at the LHC far detectors and SHiP}

\author[a]{Zeren Simon Wang\,\orcidlink{0000-0002-1483-6314}}
\emailAdd{wzs@mx.nthu.edu.tw}
\affiliation[a]{School of Physics, Hefei University of Technology, Hefei 230601, China}

\author[\,a, 1]{Yu Zhang\,\orcidlink{0000-0001-9415-8252}\note{Corresponding author.}}
\emailAdd{dayu@hfut.edu.cn}

\author[\,b, 2]{Wei Liu\,\orcidlink{0000-0002-3803-0446}\note{Corresponding author.}}
\emailAdd{wei.liu@njust.edu.cn}
\affiliation[b]{Department of Applied Physics and MIIT Key Laboratory of Semiconductor Microstructure and Quantum Sensing, Nanjing University of Science and Technology, Nanjing 210094, China}

\date{\today}

\vskip1mm
\abstract{In hidden-sector models, axion-like particles (ALPs) can couple to heavy neutral leptons (HNLs), leading to rich phenomenologies. We study ALPs produced from $D$- and $B$-meson decays via quark-flavor-violating couplings, and decaying exclusively into a pair of HNLs which mix with active neutrinos. The ALP can be either short- or long-lived, depending on the masses of the ALP and the HNL, as well as the corresponding coupling strength. Such GeV-scale HNLs are necessarily long-lived given the current bounds on their mixing parameters. We assess the sensitivities of the LHC far detectors and SHiP, to the long-lived HNLs in such theoretical scenarios. We find that for currently allowed values of the ALP couplings, most of these experiments can probe the active-sterile-neutrino mixing parameters multiple orders of magnitude beyond the present bounds, covering large parameter region targeted with the type-I seesaw mechanism. In addition, our results show that compared to the case of a promptly decaying ALP, assuming an ALP of longer lifetimes weakens the sensitivities of the considered experiments to the long-lived HNLs.}



\maketitle

%


\section{Introduction}\label{sec:intro}

Hidden-sector models postulate existence of mediator particles connecting the visible world of the Standard Model (SM) spectrum and a hidden sector containing dark matter (DM).
Notable examples include models of ``portal physics'' (see e.g.~Ref.~\cite{Antel:2023hkf} for a review) and of a hidden valley~\cite{Strassler:2006im}.
The mediator particles may emerge in distinct forms, including but not constrained to a light scalar boson that mixes with the SM-like Higgs boson~\cite{OConnell:2006rsp,Wells:2008xg,Bird:2004ts,Pospelov:2007mp,Krnjaic:2015mbs,Boiarska:2019jym}, sterile neutrinos or heavy neutral leptons (HNLs) that mix with the active neutrinos~\cite{Shrock:1980vy,Shrock:1980ct,Shrock:1981wq} (see also Ref.~\cite{Abdullahi:2022jlv} for a review), a dark photon that kinetically mixes with the SM $U(1)_Y$ gauge boson~\cite{Okun:1982xi,Galison:1983pa,Holdom:1985ag,Boehm:2003hm,Pospelov:2008zw}, and a pseudoscalar boson called axion or axion-like particle (ALP)~\cite{Peccei:1977hh,Peccei:1977ur,Witten:1984dg,Conlon:2006tq,Arkani-Hamed:2006emk,Arvanitaki:2009fg,Cicoli:2012sz}.
Such models are well motivated, providing explanations on various issues of the SM, such as the non-zero light-neutrino mass, DM, the strong CP problem, and the hierarchy problem.

In this work, we focus on the HNLs and the ALP.
The HNLs, $N$'s, are hypothetical SM-singlet fermionic fields that are under scrutiny with various kinds of theoretical and experimental efforts.
Originally proposed as sterile neutrinos in the type-I seesaw mechanism~\cite{Minkowski:1977sc,Yanagida:1979as,Mohapatra:1979ia,Gell-Mann:1979vob,Schechter:1980gr} as a way to provide mass to the light neutrinos, they are predicted to be as heavy as in the $\mathcal{O}(10^{15})$ GeV range for $\mathcal{O}(1)$ Yukawa couplings to explain the $\mathcal{O}(0.1)$-eV active-neutrino masses.
This implies extremely tiny mixings between the active neutrinos and the HNLs.
The feeble mixings and the GUT-scale masses render such hypothetical particles beyond the reach of terrestrial experiments in the near future.
However, in low-scale seesaw scenarios such as the inverse~\cite{Mohapatra:1986aw,Mohapatra:1986bd} and linear~\cite{Akhmedov:1995ip,Malinsky:2005bi} seesaw mechanisms, the mixing parameters are allowed to be larger.
Thus, much lighter, possibly GeV-scale sterile neutrinos, often called HNLs, are predicted and may be observable at current experiments.
In addition to giving rise to active-neutrino masses, the sterile neutrinos could comprise the DM if lying in the keV mass range~\cite{Dodelson:1993je,Shi:1998km,Dolgov:2000ew,Abazajian:2001vt,Abazajian:2001nj,Asaka:2005an,Asaka:2005pn,Asaka:2006nq}, and explain neutrino excesses observed in recent years at some short-baseline experiments~\cite{MiniBooNE:2022emn,Fischer:2019fbw,Bertuzzo:2018itn,Ballett:2018ynz,Bertuzzo:2018ftf}.

The ALP, $a$, is closely related to the QCD axion which was originally proposed as the pseudo-Nambu-Goldstone boson arising from the spontaneous breaking of a global $U(1)_{PQ}$ symmetry~\cite{Peccei:1977hh} in order to solve the strong CP problem.
While the axion's mass is connected to the global symmetry breaking scale, there is no such requirement on the ALP.
As a result, the mass of the ALP and its couplings to the SM particles are decoupled, leading to a multitude of possible signatures at various experiments, though the ALP may not provide a solution to the strong CP problem any more.
Furthermore, the ALP can serve as a DM candidate~\cite{Preskill:1982cy,Abbott:1982af,Dine:1982ah,Arias:2012az} and solve the hierarchy problem~\cite{Graham:2015cka}.
The ALP can couple to various fundamental particles in the SM, including leptons, quarks, gauge bosons, as well as the Higgs boson.
In this work, among these couplings we confine ourselves to those with the SM quarks, and particularly, we focus on off-diagonal couplings in either the up- or down-type quark sector and the ALP with $\mathcal{O}(0.1\text{--}1)$-GeV masses, which are strongly constrained from low-energy (meson-scale) processes~\cite{Bauer:2021mvw}.
Various UV-complete ALP models predict such off-diagonal couplings, at tree level~\cite{Ema:2016ops,Calibbi:2016hwq,Arias-Aragon:2017eww,Bjorkeroth:2018ipq,delaVega:2021ugs,DiLuzio:2023ndz,DiLuzio:2017ogq,Linster:2018avp,Alonso-Alvarez:2023wig} or loop level~\cite{Gavela:2019wzg,Bauer:2020jbp,Bauer:2021mvw,Chakraborty:2021wda,Bertholet:2021hjl}.
In addition, we assume quark-flavor-diagonal couplings to be vanishing.
Specifically, UV-complete models where the quark-flavor-diagonal couplings are suppressed include the astrophobic axion~\cite{DiLuzio:2017ogq}, where the axion in DFSZ-like models has family-dependent PQ charges thus keeping small the diagonal coupling of the axion to the first-generation quarks and also generating QFV couplings.
Further models that can realize suppression of (some) diagonal couplings and generation of off-diagonal couplings include several ones with a Froggatt-Nielsen mechanism~\cite{delaVega:2021ugs,Calibbi:2016hwq,Linster:2018avp}.
Nevertheless, in this phenomenological study, we remain agnostic about the origin of the particular ALP-quark flavor structure and treat the ALP-quark couplings as independent parameters.
For recent phenomenological studies on quark-flavor-violating (QFV) couplings of the ALP, see e.g.~Refs.~\cite{Gorbunov:2000ht,MartinCamalich:2020dfe,Carmona:2021seb,Bauer:2021mvw,Carmona:2022jid,Li:2024thq,Cheung:2024qve}.
Further, we introduce the ALP coupling to a pair of HNLs, leading to the decay $a\to N N$.
Such a coupling may provide a DM candidate or explain the observed baryon asymmetry of the Universe~\cite{Berryman:2017twh,Carvajal:2017gjj,Alves:2019xpc,Gola:2021abm,deGiorgi:2022oks,Abdullahi:2023gdj,Cataldi:2024bcs,Marcos:2024yfm,Wang:2024prt}.

We study the ALP produced in rare decays of bottom and charm mesons in association with a lighter meson (called the $B$-scenario and $D$-scenario in this work, respectively),\footnote{See also Refs.~\cite{Berezhiani:1989fs,Berezhiani:1990wn,Berezhiani:1990jj} for earlier studies on $B$-meson decays to an ALP.} and decaying into a pair of HNLs with an essentially 100\% branching fraction.
Here, the ALP production results from the QFV couplings at tree level, and its total decay width is dominantly controlled by its coupling to the HNLs for the kinematically allowed mass range.
The ALP coupling with the HNLs is constrained from perturbativity considerations~\cite{Alves:2019xpc}, and we find that under this constraint, the ALP may be either short- or long-lived\footnote{For recent studies on long-lived ALPs predicted in various theoretical scenarios, see, for instance, Refs.~\cite{Coloma:2023oxx,Afik:2023mhj,Cheung:2021mol,Wang:2024zky,Carmona:2022jid,Cheung:2024qve}.}, depending on the coupling strength and the masses of the ALP and the HNL. 
The HNLs then decay via weak interactions, allowed by their mixings with the active neutrinos, into SM particles.
The HNLs in the considered mass range are typically long-lived, given the current bounds on these mixing parameters.
We therefore propose to search for such HNLs at present and future experiments dedicated to long-lived-particle (LLP) searches (see Refs.~\cite{Alimena:2019zri,Lee:2018pag,Curtin:2018mvb,Beacham:2019nyx} for some reviews on LLP).
Indeed, HNLs in this mass range have been intensively searched for mainly at beam-dump and fixed-target experiments, and stringent bounds have been therefrom obtained primarily in the ``minimal'' scenarios where the HNLs only participate in the SM weak interactions via the active-sterile-neutrino mixings~\cite{PIENU:2017wbj,Bryman:2019bjg,CHARM:1985nku,NA62:2020mcv,T2K:2019jwa,Barouki:2022bkt}.
In these minimal scenarios, while the model is simple, both the production and decay are mediated via the active-sterile-neutrino mixings.
This typically results in relatively inferior sensitivities compared to those expected in theoretical scenarios where the production and decay proceed via different couplings.
The scenario we study is one such example, and we therefore anticipate much stronger sensitivities.

When comparing the sensitivity reach we derive in the ALP-HNL model with the existing bounds obtained in the minimal scenarios of the HNLs, we should check if the existing bounds apply to our model.
In particular, for the $D$- and $B$-scenarios we consider, only the bounds attained with considering $D$- and $B$-meson decays, respectively, could be affected.
As we will explain later, in this work only in the $D$-scenario, the leading bounds should be reinterpreted in terms of our ALP-HNL model.
These bounds were derived in Ref.~\cite{Barouki:2022bkt} and Ref.~\cite{Boiarska:2021yho} which re-analyzed the original results obtained from searches at BEBC~\cite{WA66:1985mfx} and CHARM~\cite{CHARM:1985nku}, respectively, where the HNLs emanate from $D$-meson decays.
Instead of performing a full recast and reinterpretation, here, we follow a simple method proposed in Ref.~\cite{Beltran:2023nli} re-scaling the HNL production and decay rates, in order to derive the corresponding reinterpreted bounds on our $D$-scenario.\footnote{The kinematics between the processes we study (HNLs produced from an on-shell ALP which is produced in $D$-meson decays) and those targeted in the experimental searches (HNLs produced directly from $D$-meson decays) differ. In principle, this should make the usage of the simple recast method invalid. However, for reasons of the strong forward boost at the proton beam-dump experiments, we expect the difference is minor in the case of a short-lived ALP and therefore still choose to follow the procedure. For the case of a long-lived ALP, we will display the same bounds, admitting a compromised accuracy because of the ALP's macroscopic decay distance.}

We focus on ongoing and proposed LLP-search experiments at CERN, all designed with the capability to reconstruct the displaced vertices (DVs) of tracks arising from the long decays of the LLPs.
The first class consists of a series of ``far detectors'' with a distance of about 5 - 600 meters to various interactions points (IPs) at the LHC.
This includes ANUBIS~\cite{Bauer:2019vqk}, CODEX-b~\cite{Gligorov:2017nwh,Aielli:2019ivi}, FACET~\cite{Cerci:2021nlb}, FASER and FASER2~\cite{Feng:2017uoz,FASER:2018eoc}, MoEDAL-MAPP1 and MAPP2~\cite{Pinfold:2019nqj,Pinfold:2019zwp}, as well as MATHUSLA~\cite{Curtin:2018mvb,MATHUSLA:2020uve,Chou:2016lxi}.
As a huge number of mesons are produced at the LHC IPs with a large boost in the forward direction, these experiments typically have particularly good acceptance to the LLPs produced from meson decays.
Further, with a macroscopic distance from the IP, most of these experiments are protected by various shielding measures, thus suffering from little or no background, and are therefore particularly powerful for LLP searches.
Studies on their sensitivity reaches to long-lived HNLs in the minimal scenarios can be found in Refs.~\cite{Curtin:2018mvb,Aielli:2019ivi,DeVries:2020jbs,Kling:2018wct,Helo:2018qej,Hirsch:2020klk,Ovchynnikov:2022its,FASER:2018eoc}.
In addition, we include the recently approved experiment, Search for Hidden Particles (SHiP)~\cite{SHiP:2015vad,Alekhin:2015byh,SHiP:2018xqw,SHiP:2021nfo,Albanese:2878604}, in the study.
At SHiP, a proton beam of 400 GeV energy extracted from the SPS accelerator at CERN impinges on a high-density proton target, resulting in an enormous number of pseudoscalar mesons produced.
In particular, at SHiP, a Hidden Sector Decay Spectrometer (HSDS) with a large fiducial decay volume is planned to be implemented.
It is specifically devoted to searches for feebly interacting new physics, and has the capabilities of DV reconstruction, invariant-mass measurement, and particle identification of the LLP decay products, in a background-free environment~\cite{Albanese:2878604}.
Some recent phenomenological research works on searches for long-lived HNLs at SHiP can be found in Refs.~\cite{SHiP:2018xqw,Gorbunov:2020rjx,DeVries:2020jbs,Ipek:2023jdp,deVries:2024mla}.

This work is organized as follows.
In Sec.~\ref{sec:models} we detail the ALP-HNL model, followed by Sec.~\ref{sec:experiment} dedicated to introducing the considered experimental setups, the procedure of our numerical simulation, as well as the computation of the signal-event rates.
We then present the numerical results in Sec.~\ref{sec:results} and conclude the work with a summary and an outlook in Sec.~\ref{sec:conclusions}.

\section{Theoretical model}\label{sec:models}

We consider a low-energy effective Lagrangian for the ALP up to dimension 5 with the following terms~\cite{Bauer:2017ris,Bauer:2018uxu,Bauer:2021mvw,Carmona:2021seb,Beltran:2023nli},
\begin{eqnarray}
\mathcal{L}_{a} = \frac{1}{2} \, \partial_\mu a  \,\partial^\mu a - \frac{1}{2}m_a^2 a^2  + \frac{\partial_\mu a}{\Lambda}  \sum_q \sum_{i,j} g^q_{i,j} \bar{q_i}\gamma^\mu q_j,
\label{eqn:Lag_ALP}
\end{eqnarray}
where $m_a$ labels the ALP mass, $\Lambda$ is the effective cut-off scale, and $g^q_{i,j}$ are dimensionless couplings with $q$ going over $u_L, u_R, d_L,$ and $d_R$.
In particular, we will study two benchmark scenarios of the flavor indices $(i, j)$, namely, $(2, 1)$ for up-type quarks and $(3, 2)$ for down-type quarks, leading to $c\to u$ and $b\to s$ transitions at the quark level, respectively.
Such transitions would emerge at the hadron level at experiments, in terms of charm and bottom mesons' decays, respectively.
Specifically, we focus on $P\to P'/V a$ decays in this work for on-shell production of the ALP $a$, where $P$ and $P'$ denote pseudoscalar mesons and $V$ labels a vector meson.
As mentioned in Sec.~\ref{sec:intro}, these two scenarios are referred to as the $D$- and $B$-scenarios, respectively.

For $g^u_{2,1}=g^{u_R}_{2,1}+g^{u_L}_{2,1}$, the following $D$-mesons' transitions are induced: $D^0\to \pi^0/\eta/\eta'$, $D^+\to \pi^+$, and $D_s^+\to K^+$, in association with an ALP $a$.
The $D$-mesons can also decay to a vector meson plus an ALP, with the coupling $g^u_{2,1}=g^{u_R}_{2,1}-g^{u_L}_{2,1}$: $D^0\to \rho^0/\omega \,a$, $D^+\to \rho^+ \,a$, and $D_s^+\to K^{*+} \,a$.
The $B$-mesons decay into a lighter meson and an ALP, when we consider the couplings $g^d_{3,2}=g_{3,2}^{d_R}+g_{3,2}^{d_L}$ and $g^d_{3,2}=g_{3,2}^{d_R}-g_{3,2}^{d_L}$.
For the former coupling, we have the $B^0\to K^0, B^+\to K^+,$ and $B_s^0\to \eta/\eta'$ transitions, and for the latter the following ones can take place: $B^0\to K^{*0}, B^+\to K^{*+}$, and $B^0_s\to \phi$.
In this work, we assume that either $g_{i,j}^{q_L}$ or $g_{i,j}^{q_R}$ is zero, so that both $P\to P'$ and $P\to V$ transitions are mediated by a single coupling, and we note that the charge-conjugated channels of the above-mentioned ones are taken into account in the numerical study.

The ALP production rates in these meson decays induced by the coupling $g^q_{i,j}$ are computed with the following expressions~\cite{Bauer:2021mvw}:
\begin{eqnarray}
 \Gamma\left(P \to P' a\right) &=& f\, \frac{|g^q_{i,j}|^2}{64\pi\Lambda^2}
 \left|F_0^{P \to P'}(m_a^2)\right|^2 m_P^3 \left(1 - \frac{m_{P'}^2}{m_P^2}\right)^2
 \lambda^{1/2}\left(\frac{m_{P'}^2}{m_P^2}, \frac{m_a^2}{m_P^2}\right)\,,\label{eqn:GammaP2Pprimea} \\
 \Gamma\left(P \to V a\right) &=& h\, \frac{|g^q_{i,j}|^2}{64\pi\Lambda^2} 
 \left|A_0^{P \to V}(m_a^2)\right|^2 m_P^3 \,  \lambda^{3/2}\left(\frac{m_{V}^2}{m_P^2}, \frac{m_a^2}{m_P^2}\right)\,, \label{eqn:GammaP2Va}
\end{eqnarray}
where $\lambda(x,y)\equiv 1+x^2+y^2-2\,x -2\,y -2xy$ and $m_{P/P'/V}$ denotes the mass of the $P/P'/V$ meson.
$F_0^{P\to P'}$ and $A_0^{P\to V}$ are transition form factors~\cite{Wirbel:1985ji}.
$f=h=1$ except in the following neutral-meson transitions:
\begin{eqnarray}
    &f=1/2\text{ for }D^0\to \pi^0, &f=2/3\text{ for }D^0/B_s^0 \to \eta,\nonumber \\
    &f=1/3\text{ for }D^0/B_s^0\to \eta', &h=1/2\text{ for }D^0\to \rho^0/\omega.
\end{eqnarray}

The data of the relevant transition form factors are extracted from the literature.
For the $D\to \pi$ transition form factors, Ref.~\cite{Lubicz:2017syv} is used, and for the other $D\to P'$ and for all the $D\to V$ transitions the form factors are provided in Ref.~\cite{Ivanov:2019nqd}.
Further, the $B\to K$ and $B_s^0\to \eta/\eta'$ form factors are taken from Ref.~\cite{FlavourLatticeAveragingGroupFLAG:2021npn} and Ref.~\cite{Wu:2006rd}, respectively.
Finally, $A_0^{B\to K^*}(m_a^2)$ and $A_0^{B_s^0\to \phi}(m_a^2)$ are extracted from Ref.~\cite{Bharucha:2015bzk}.
See also Ref.~\cite{Beltran:2022ast} for more detail on these form factors.

Here, we note that Ref.~\cite{DallaValleGarcia:2023xhh} pointed out the importance of taking into account additional kaon resonances including scalar, axial-vector, tensor, as well as further vector ones (see also Ref.~\cite{Boiarska:2019jym}) that could enhance the production rates of the ALP in $B\to K a$ decays by a factor of about 4.
In the numerical analysis, instead of computing the ALP production rates in these additional channels, we will simply re-scale those of the $B \to K a$ channels listed above by 4.

The present bounds on $g^u_{2,1}/\Lambda$ and $g^d_{3,2}/\Lambda$ can be derived from the existing limits on the decay branching ratios (Br) of $D^0\to \pi^0 \nu\bar{\nu}$ and $B\to K\nu \bar{\nu}$, since the ALP decays exclusively to a pair of long-lived HNLs.
For the $D^0\to \pi^0 \nu\bar{\nu}$ channel, we assume that the difference in the kinematics between the two-body decay $D^0\to \pi^0 a$ considered in this work and the three-body decay targeted in the experimental search would not lead to large deviations in the results; using Eq.~\eqref{eqn:GammaP2Pprimea} we find $g^u_{2,1}/\Lambda\lesssim 2\times 10^{-4}$ TeV$^{-1}$ (derived also in Ref.~\cite{Beltran:2023nli}) based on the present upper bound obtained at BESIII~\cite{BESIII:2021slf}.
In numerical analysis, we will choose benchmark values of $m_a$ at 0.5, 1.0, and 1.5 GeV, and will fix the value of $g^u_{2,1}/\Lambda$ at its upper bound, $2\times 10^{-4}$ TeV$^{-1}$.
For the $B$-scenario, we will consider $m_a$ at 1.0, 2.5, and 4.0 GeV, and assume, for these ALP masses, values of $g^d_{3,2}/\Lambda$ that are allowed by the latest measurement of Br$(B^+\to  K^+ \nu\bar{\nu})_{\text{exp}}$ at Belle II~\cite{Belle-II:2023esi} and the present upper bounds on Br$(B^{+/0}\to K^{*+/0} \nu\bar{\nu})$~\cite{Belle:2013tnz,Belle:2017oht,ParticleDataGroup:2024cfk}.
The recent Belle II measurement reports Br$(B^+\to  K^+ \nu\bar{\nu})_{\text{exp}}=(2.3\pm 0.7)\times 10^{-5}$, which is $2.7\sigma$ above the SM prediction, Br$(B^+\to K^+\nu  \bar{\nu})_{\text{SM}}^{{\text{SD}}}=({4.43\pm 0.31})\times 10^{-6}$~\cite{Becirevic:2023aov} {where the superscript ``SD'' stands for ``short-distance'' implying that tree-level long-distance contributions~\cite{Kamenik:2009kc} have been subtracted, in accordance with the Belle II search report~\cite{Belle-II:2023esi} where such long-distance contributions are included as part of background}.
This implies a new-physics (NP) window, Br$(B^+\to K^+\nu\bar{\nu})_{\text{NP}}=(1.9\pm 0.7)\times 10^{-5}$~\cite{He:2023bnk}, for accommodating the excess.
Values of Br$(B^+\to K^+\nu\bar{\nu})_{\text{NP}}$ outside this window cannot explain the excess, and values above (below) the window are excluded (allowed).
Moreover, the distribution of the invariant mass squared of the missing energy ($q^2$) is given in the Belle II search report~\cite{Belle-II:2023esi}.
Here, instead of performing a detailed fit analysis on the $q^2$ bins, we choose to take $g^d_{3,2}/\Lambda=4 \times 10^{-6}$ TeV$^{-1}$ for both $m_a=1.0$ and 2.5 GeV, and $g^d_{3,2}/\Lambda=1 \times 10^{-6}$ TeV$^{-1}$ for $m_a=4.0$ GeV, which correspond to Br$(B^+\to K^+ a)\sim 4.0\times 10^{-6}$ and $2.4\times 10^{-7}$, respectively.
These values of Br$(B^+\to K^+ a)$ are below the $2\sigma$ window of Br$(B^+\to K^+\nu\bar{\nu})_{\text{NP}}=(1.9\pm 0.7)\times 10^{-5}$ quoted above.
Finally, in Ref.~\cite{He:2022ljo} the NP bounds to accommodate the present upper limits on Br$(B^+\to K^{*+} \nu \bar{\nu})\lesssim 4.0\times 10^{-5}$~\cite{Belle:2013tnz} and Br$(B^0\to K^{*0}\nu \bar{\nu})\lesssim 1.8\times 10^{-5}$~\cite{Belle:2017oht} are determined to be $3.1\times 10^{-5}$ and $1.0\times 10^{-5}$, respectively.
We have checked and verified with Eq.~\eqref{eqn:GammaP2Va} that our chosen values of $g^d_{3,2}/\Lambda$ with the corresponding ALP masses to mediate the decays $B^{+/0}\to K^{*+/0} a$ are all allowed by these bounds.

Now turning to the HNLs, we first comment that we will consider the case that there is only one kinematically relevant HNL $N$.
We present in the following the Lagrangian for the HNL interactions with the SM electroweak gauge bosons via charged and neutral currents:
\begin{eqnarray}
		\mathcal{L}_{N} &=& \frac{g}{\sqrt{2}}\ \sum_{\alpha}
		V_{\alpha N}\ \bar \ell_\alpha \gamma^{\mu} P_L N W^-_{L \mu}
		+		\frac{g}{2 \cos\theta_W}\ \sum_{\alpha, i}V^{L}_{\alpha i} V_{\alpha N}^* \overline{N} \gamma^{\mu} P_L \nu_{i} Z_{\mu},
		\label{eqn:Lag_HNL}
\end{eqnarray}
where $g$ is the SU(2) gauge coupling and $\theta_W$ is the weak mixing angle.
$V_{\alpha N}$ is the mixing angle between the HNL and the active neutrino $\nu_\alpha$ with $\alpha=e, \mu, \tau$, and $V^L$ is the neutrino mixing matrix in the left-handed sector.
Further, $\nu_i$ denotes the active-neutrino mass eigenstates with $i=1, 2, 3$.
For the purpose of a simple phenomenological study, we assume a Majorana HNL $N$ and consider the benchmark scenario where the HNL mixes with the electron neutrino only.
Therefore, in the end, we have two parameters within the HNL sector, $m_N$ (the mass of the HNL) and $|V_{eN}|^2$, at play, which we assume as independent parameters.
The two-body and three-body decay widths of the HNL at tree level are then calculated with formulas given in Refs.~\cite{Atre:2009rg,DeVries:2020jbs,Bondarenko:2018ptm}.

The effective Lagrangian describing the interaction between the ALP $a$ and the HNL $N$ is given below~\cite{Alves:2019xpc,Gola:2021abm,Marcos:2024yfm}:
\begin{eqnarray}
    \mathcal{L}_{a, N} =   \frac{\partial_\mu a}{\,\Lambda}\, g_N   \overline{N}  \gamma^\mu \gamma_5  N,
    \label{eqn:Lag_ALP_HNL}
\end{eqnarray}
where $g_N$ is a dimensionless coupling. 
We first follow Ref.~\cite{Alves:2019xpc} to fix $g_N/\Lambda=10^{-3}$ GeV$^{-1}$, ensuring perturbativity be obeyed with $\frac{g_N}{\Lambda}m_N<1$ (even though for our considered $m_N$ range, choosing $g_N/\Lambda$ of up to about $10^{-1}$ GeV$^{-1}$ would still be allowed).
A smaller value of $g_N/\Lambda$, at $10^{-8}$ GeV$^{-1}$, is additionally chosen for numerical study, with which the ALP becomes long-lived.
The ALP in the theoretical scenarios considered here decays exclusively to a pair of $N$'s, via the $g_N$ coupling, and the corresponding decay width is given below~\cite{Alves:2019xpc}:
\begin{eqnarray}
    \Gamma(a\to N N)=\frac{1}{2\pi} \Big(\frac{g_N}{\Lambda}\Big)^2  m_N^2 \, m_a \sqrt{1-\frac{4 m_N^2}{m_a^2}}.\label{eqn:GammaALP2NN}
\end{eqnarray}
\begin{figure}[t]
	\centering
	\includegraphics[width=0.7\textwidth]{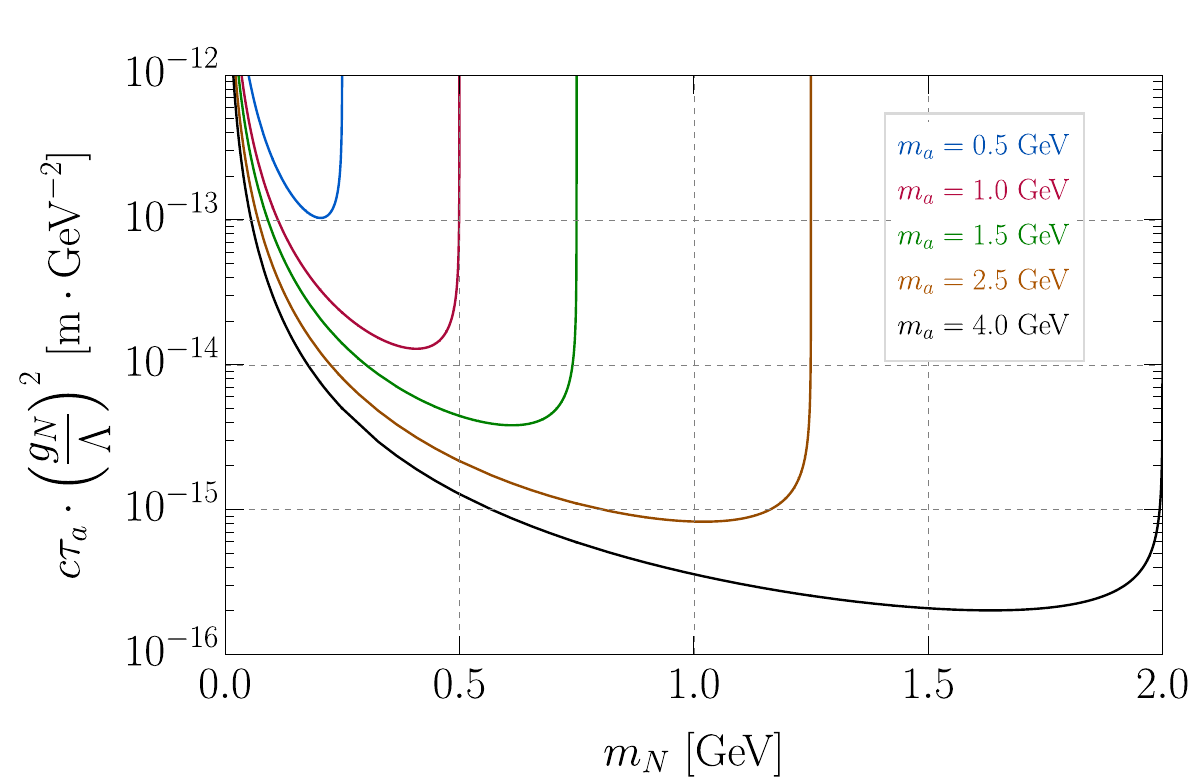}
	\caption{$c\tau_a\cdot\Big( \frac{g_N}{\Lambda}\Big)^2$ vs.~$m_N$ for various choices of $m_a$.
 }
 \label{fig:ctau_ALP_vs_mN}
\end{figure}
In Fig.~\ref{fig:ctau_ALP_vs_mN}, we present a plot of $c\tau_a\cdot\Big( \frac{g_N}{\Lambda}\Big)^2$ as a function of $m_N$, for the values of $m_a$ which we will use for numerical analysis.
Here, $c\tau_a$ is saturated by the decay $a\to N N$ and is thus derived from Eq.~\eqref{eqn:GammaALP2NN} solely.
The ALP is shown to have the largest lifetimes at small values of $m_a$ or $m_N$ and at the vicinity of the kinematic threshold ($m_N\lesssim m_a/2$), cf.~Eq.~\eqref{eqn:GammaALP2NN}.

\begin{figure}[t]
	\centering
	\includegraphics[width=0.7\textwidth]{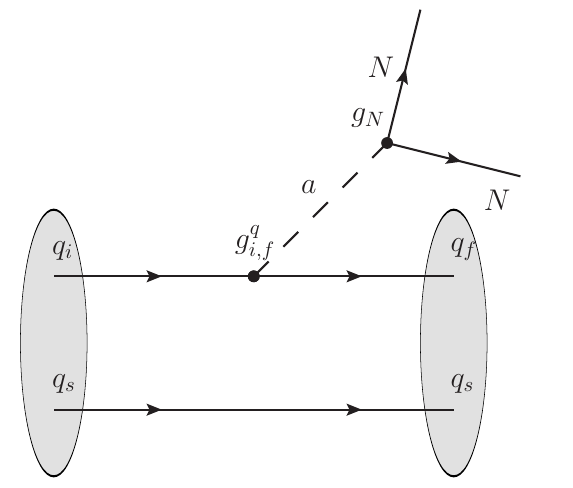}
	\caption{Feynman diagram depicting the HNL-production signal process. Here, the coupling $g^q_{i,f}$ mediates the quark-level transition between the initial and final quarks $q_i \to q_f$ in association with an ALP, and $q_s$ denotes a spectator quark.
    The initial and final mesons consist of the quarks $(q_i, q_s)$ and $(q_f, q_s)$, respectively.
    The ALP $a$ decays into a pair of HNLs via $g_N$, either promptly or after traveling a macroscopic distance.
 }
 \label{fig:feynman_ALP_production}
\end{figure}
In the end of this section, we show the Feynman diagram for the production process of the HNLs in Fig.~\ref{fig:feynman_ALP_production}.
Here, the initial and final mesons are composed of the quarks $(q_i, q_s)$ and $(q_f, q_s)$, respectively.
The ALP $a$ is produced via the coupling $g_{i,f}^q$ and decays to a pair of $N$'s via $g_{N}$.

\section{Experimental setups and simulation procedure}\label{sec:experiment}

At the LHC, a series of ``far-detector'' experiments have been proposed, approved, or even in operation, dedicated to searches for LLPs predicted in various new-physics models.
By reconstructing the DVs consisting of tracks inside the detector's fiducial volume, these experiments can detect LLP decays that are hardly accessible at the LHC main detectors such as ATLAS and CMS.
Concretely, in this work, we consider the experiments of ANUBIS~\cite{Bauer:2019vqk}, CODEX-b~\cite{Gligorov:2017nwh,Aielli:2019ivi}, FACET~\cite{Cerci:2021nlb}, FASER and FASER2~\cite{Feng:2017uoz,FASER:2018eoc}, MoEDAL-MAPP1 and MAPP2~\cite{Pinfold:2019nqj,Pinfold:2019zwp}, and MATHUSLA~\cite{Curtin:2018mvb,MATHUSLA:2020uve,Chou:2016lxi}.

We briefly introduce them according to their associated IP separately.
ANUBIS, FASER, and FASER2 are located in the vicinity of the ATLAS IP.
ANUBIS is supposed to be installed in a service shaft above the ATLAS IP with an $\mathcal{O}(1)$ m, slight horizontal displacement along the $z$-axis.\footnote{A new design of the ANUBIS experiment has been proposed~\cite{ANUBIS_talk_slides}, suggesting to place ANUBIS at the ATLAS cavern ceiling or shaft bottom instead. At this changing stage of ANUBIS, however, we decide to stick with the original proposal in the present sensitivity study.}
It should have a radius of 9 m and a height of 54 m, and is slated to collect data of up to 3000 fb$^{-1}$ integrated luminosity during the high-luminosity LHC (HL-LHC) phase.
FASER has been implemented and is currently collecting data during the LHC Run 3.
The FASER collaboration has published its initial results~\cite{FASER:2023zcr,FASER:2023tle,FASER:2024hoe}.
It is a small cylindrical detector with a length of 1.5 m and a radius of 0.1 m, placed in the very forward direction of the IP with a distance of about 480 m along the $z$-axis.
A total of 150 fb$^{-1}$ integrated luminosity is expected to be collected.
FASER2 would be an upgraded version of FASER, with a radius of 1 m, a length of 5 m, and a distance of 620 m from the ATLAS IP, planned to collect 3000 fb$^{-1}$ of data in the HL-LHC period.

At the CMS IP, FACET and MATHUSLA have been proposed for LLP searches, expected to receive 3000 fb$^{-1}$ data during the HL-LHC era.
FACET is to be constructed in the very forward direction of the IP along the $z$-axis.
With a distance of 101 m from its near end to the IP, FACET is to be installed in the beam-pipe region, thus designed to have an inner (outer) radius of 18 cm (50 cm), and a length of 18 m.
MATHUSLA is unique in the sense that it is the only proposal among the considered experiments that is to be instrumented on the ground surface.
For its geometrical design, we follow Ref.~\cite{MATHUSLA:2020uve}, considering a $100\text{ m}\times 100 \text{ m}\times 25\text{ m}$ box shape.
Parallel with the beam axis, its near side is displaced along the beam axis (vertical axis) by 68 m (60 m)\footnote{For cost considerations, a somewhat smaller version of MATHUSLA has been proposed recently~\cite{mathusla_new_design}. As it is not yet finalized, in this work we stick to the design discussed in Ref.~\cite{MATHUSLA:2020uve}.}.

CODEX-b, MoEDAL-MAPP1, and MoEDAL-MAPP2 are experiments associated with the LHCb IP.
CODEX-b is a proposed box-shaped detector, with a $10\text{ m}\times 10\text{ m}\times 10\text{ m}$ dimension and a distance of 25 m from the IP, covering $\eta\in [0.2, 0.6]$ and $\frac{\delta\phi}{2\pi}\sim \frac{0.4}{2\pi}$ where $\eta$ labels the pseudorapidity and $\phi$ the azimuthal angle.
MoEDAL-MAPP1 and MAPP2 are trapezoidal-shaped detectors in the UGCI gallery near the LHCb IP, and have a volume of 130 $\text{m}^3$ and 430 $\text{m}^3$, respectively.
MAPP1 is at a polar angle of 5$^\circ$ with respect to the IP with a distance of 55 m, and MAPP2 should occupy the whole gallery.
MoEDAL-MAPP1 is expected to receive 30 fb$^{-1}$ integrated luminosity data during the LHC Run3 while both CODEX-b and MoEDAL-MAPP2 are planned to collect 300 fb$^{-1}$ data in the HL-LHC phase.

SHiP~\cite{SHiP:2015vad,Alekhin:2015byh,SHiP:2018xqw,SHiP:2021nfo,Albanese:2878604} is an approved beam-dump experiment, planned to start operation in 2031.
At SHiP, a 400 GeV proton beam extracted from the CERN SPS accelerator hits a heavy proton target, leading to large production rates of mesons.
The SHiP experiment consists of an upstream system of the Scattering and Neutrino Detector, and a downstream one of the HSDS.
In this work, we focus on the search for long-lived HNLs and hence on the HSDS apparatus.
We follow Ref.~\cite{Albanese:2878604} for the latest design of the HSDS.
The detector is situated $\sim 33$ m downstream from the target, and has a pyramidal-frustum-shaped fiducial decay volume.
The decay volume is 50 m long, and has a front (rear) surface of dimensions 1.0 m $\times$ 2.7 m (4.0 m $\times$  6.0 m) in width times height.
The SPS can deliver up to $4\times 10^{19}$ protons on target (POTs) per year for 15 years, with an operation duration of 200 days per year, thus in total accumulating $6\times 10^{22}$ POTs.

Most of the LHC far detectors are equipped with sufficient shielding between the IPs and the detector, allowing for a background-free environment.
Two exceptions are ANUBIS and FACET, where certain levels of background are expected.
Here, we follow the usual practice in the literature, assuming optimistically as well zero background at these experiments.
As for SHiP, the implementation of an upstream background tagger, an surrounding background tagger, a timing detector, etc., also ensures less than one background event during the entire 15-year operation.
Therefore, in this work, we will assume zero background for all the considered experiments, and show sensitivity curves corresponding to 3 signal events and hence exclusion bounds at 95\% confidence level (C.L.).

\begin{table}[]
\resizebox{\textwidth}{!}{
\begin{tabular}{cc|cc|cc}
\multicolumn{2}{c|}{$N_{\overset{(-)}{D^0}}$} & \multicolumn{2}{c|}{$N_{D^\pm}$} & \multicolumn{2}{c}{$N_{D_s^\pm}$} \\ \hline
HL-LHC                      & SHiP       & HL-LHC             & SHiP       & HL-LHC                  & SHiP    \\
                    $3.89\times 10^{16}$                  &      $1.29\times 10^{18}$            &    $2.04\times 10^{16}$             & $4.2\times 10^{17}$              &    $6.62\times 10^{15}$              & $1.8\times 10^{17}$            \\ \hline
\multicolumn{2}{c|}{$N_{\overset{(-)}{B^0}}$}                   & \multicolumn{2}{c|}{$N_{B^\pm}$}                    & \multicolumn{2}{c}{$N_{\overset{(-)}{B_s^0}}$}           \\     \hline
HL-LHC                      & SHiP       & HL-LHC             & SHiP       & HL-LHC                  & SHiP    \\
                 $1.46\times 10^{15}$            &        $8.1\times 10^{13}$            &     $1.46\times 10^{15}$            &  $8.1\times 10^{13}$             &   $2.53\times 10^{14}$                   &     $2.16\times 10^{13}$           
\end{tabular}
}
\caption{Summary of the total numbers of the relevant $B$- and $D$-mesons estimated to be produced at the HL-LHC and SHiP, reproduced from Refs.~\cite{DeVries:2020jbs,Bondarenko:2018ptm}.
For the HL-LHC, an integrated luminosity of 3 ab$^{-1}$ is taken, and for the SHiP experiment, we assume a nominal operation of 15 years, yielding $6\times 10^{20}$ POTs.
All the numbers listed here are for a solid angle of $4\pi$, and for the sum of the numbers of both charge-conjugated mesons (such as $D^0$ and $\overline{D^0}$).}
\label{tab:meson_numbers}
\end{table}

In Table~\ref{tab:meson_numbers}, we list the total numbers of $B$- and $D$-mesons relevant to our study estimated to be produced at the HL-LHC and SHiP, extracted from Refs.~\cite{DeVries:2020jbs,Bondarenko:2018ptm}.

We perform Monte-Carlo (MC) simulations in order to determine the acceptance of the considered experiments to the long-lived HNLs.
For the LHC far detectors, we make use of Displaced Decay Counter (DDC)~\cite{Domingo:2023dew} which loads Pythia8.3~\cite{Bierlich:2022pfr} for event generation at the center-of-mass energy of 14 TeV of back-to-back proton-proton collisions and then computes the acceptances.
Concretely, we apply the \textit{HardQCD:hardbbbar} and \textit{HardQCD:hardccbar} modules of Pythia for the bottom- and charm-mesons' generation in the two benchmark scenarios, separately.
For both theoretical benchmark scenarios (the $B$- and $D$-scenarios) studied here, we choose a few benchmark values of $m_a$, and scan over the mass and proper decay length of the HNL.
As currently the SHiP detector is not yet implemented in DDC, we resort to the corner-point method applied in Ref.~\cite{Dreiner:2020qbi} for the modelling of the SHiP HSDS.
In detail, we implement a program which first loads Pythia8 to call the \textit{HardQCD:hardbbbar} or \textit{HardQCD:hardccbar} modules for the event generation.
Then, we provide the coordinates of the eight corner points of the pyramidal-frustum-shaped fiducial volume of the HSDS for the program which then determines if the path of a simulated HNL passes through the decay volume and, if so, computes the probability of it decaying inside the fiducial volume taking in the kinematic information of the HNL extracted from Pythia8.
For these events, we set the energy of the first proton beam at 400 GeV, and that of the second one at the proton rest mass, in order to simulate the beam-dump interactions at SHiP.
Furthermore, the above-mentioned tools are applicable only for the HNLs produced at the IP, i.e.~the ALP is promptly decaying.
For the case of a long-lived ALP, we implement a code inspired by Ref.~\cite{Gunther:2023vmz} which takes into account the (displaced) decay position of the simulated ALP and the moving direction of the HNL, and computes, with Pythia8 simulations, the acceptance of the studied experiments to the long-lived HNLs.

With the acceptance $\epsilon_N$ at a detector determined with MC simulations, we proceed to compute the signal-event rates with the following formula:
\begin{eqnarray}
    N_S = \sum_P N_P \cdot \text{Br}(P\to P'/V \, a)\cdot \text{Br}(a\to N N) \cdot 2 \cdot \epsilon_N \cdot \text{Br}(N\to \text{vis.}), \label{eqn:NS}
\end{eqnarray}
where $P$ denotes the pseudoscalar mesons considered in either the $D$-scenario ($D^0$, $D^+$, $D_s^+$, and their charge-conjugated states) or the $B$-scenario ($B^0$, $B^+$, $B^0_s$, and their charge-conjugated states),  Br$(a\to N N)=100\%$, and the factor 2 accounts for the fact that there are two $N$'s in each signal event.
Br$(N\to \text{vis.})$ is the decay branching ratio of the HNL into visible final states, for which we include all the final states except the tri-neutrino channels.
We assume 100\% detector efficiency for the considered experiments.
The tools we use for estimating $\epsilon_N$ take the arithmetic average of the decay probability of all the simulated HNLs:
\begin{eqnarray}
    \epsilon_N = \frac{1}{2\, N_{\text{MC}}} \sum_{i=1}^{2\, N_{\text{MC}}} \epsilon_i,
\end{eqnarray}
where $\epsilon_i$ is the individual decay probability of the $i^{\text{th}}$ simulated $N$ in a detector, $N_{\text{MC}}$ is the number of MC-simulated events, and the factor 2 is due to the same reason stated above for the factor 2 in Eq.~\eqref{eqn:NS}.
We note that $i$ is counted from $1$ to $2\,N_{\text{MC}}$.
$\epsilon_i$ is computed by considering the exponential distribution of the decay position of the long-lived HNL, with input of the production position, three-momentum, mass, and proper decay length of the HNL, as well as the detector's geometry and position with respect to the IP.
We refer to Refs.~\cite{Domingo:2023dew,Dreiner:2020qbi,DeVries:2020jbs,Gunther:2023vmz,deVries:2024mla} and the references therein for more detail.

\section{Numerical results}\label{sec:results}

\begin{figure}[t]
	\centering
	\includegraphics[width=0.495\textwidth]{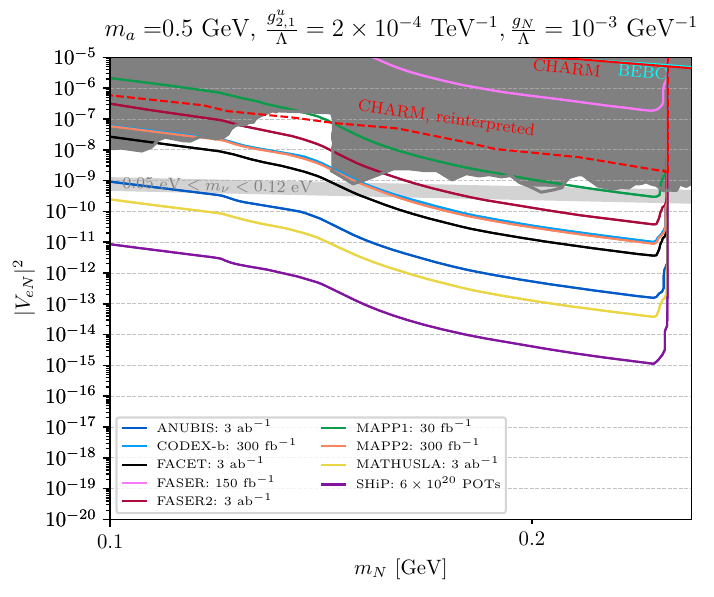}
	\includegraphics[width=0.495\textwidth]{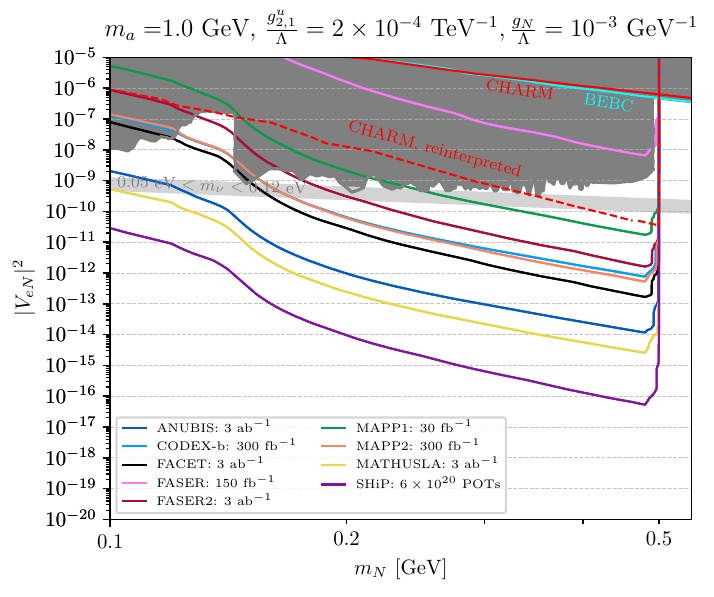}
	\includegraphics[width=0.495\textwidth]{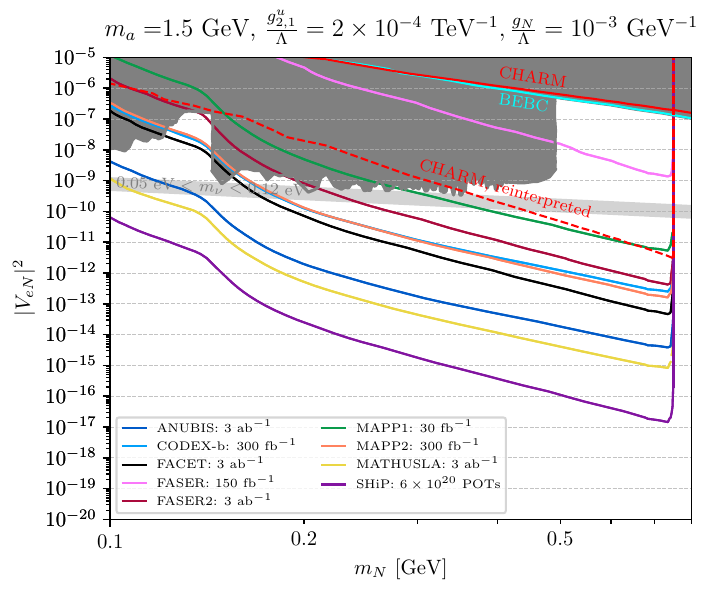}
	\caption{Sensitivity reach of the LHC far detectors and SHiP for the $D$-scenario, shown in the plane $|V_{eN}|^2$ vs.~$m_N$, for $m_a=0.5, 1.0,$ and $1.5$ GeV.
    We have fixed $\frac{g^u_{2,1}}{\Lambda}=2\times 10^{-4}$ TeV$^{-1}$ and $\frac{g_N}{\Lambda}=10^{-3}$ GeV$^{-1}$, obeying experimental measurement and perturbativity requirement, respectively. The dark gray area is the parameter regions currently excluded from fixed-target and beam-dump experiments ~\cite{PIENU:2017wbj,Bryman:2019bjg,CHARM:1985nku,NA62:2020mcv,T2K:2019jwa,Barouki:2022bkt} for the minimal scenario. These bounds are also valid for our ALP-HNL model, except those from CHARM~\cite{Boiarska:2021yho} and BEBC~\cite{Barouki:2022bkt}, which are re-drawn with red and cyan solid lines, respectively;  they are specifically shown here because they should be reinterpreted in terms of our ALP-HNL model.  Correspondingly, the red dashed curves are the reinterpreted CHARM bounds we obtain with a simple recast method considering re-scaling the production and decay rates of the HNLs~\cite{Beltran:2023nli}.  The  light gray band corresponds to the region targeted by the type-I seesaw mechanism for the active-neutrino mass between 0.05 eV and 0.12 eV.
 }
 \label{fig:sensitivity_D_1em3}
\end{figure}

\begin{figure}[t]
	\centering
	\includegraphics[width=0.495\textwidth]{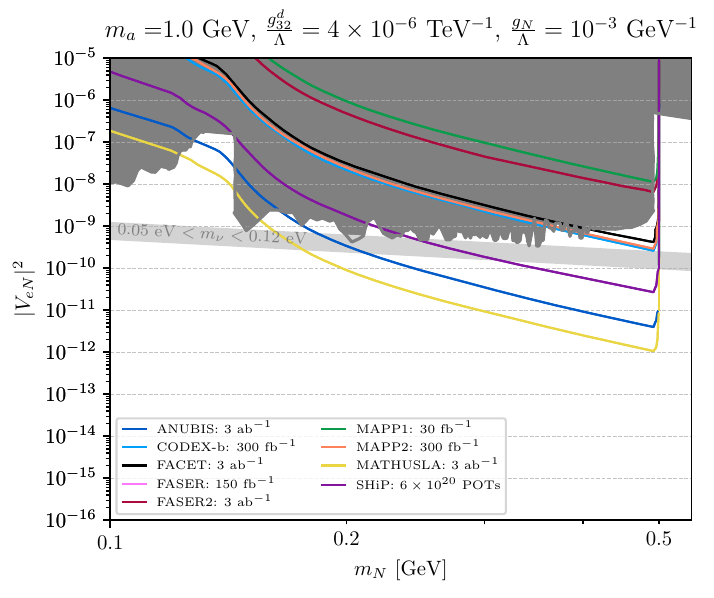}
	\includegraphics[width=0.495\textwidth]{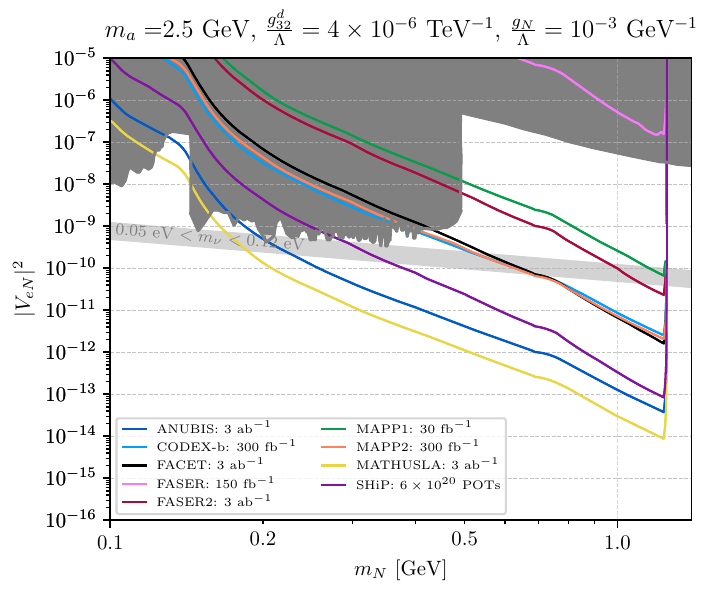}
	\includegraphics[width=0.495\textwidth]{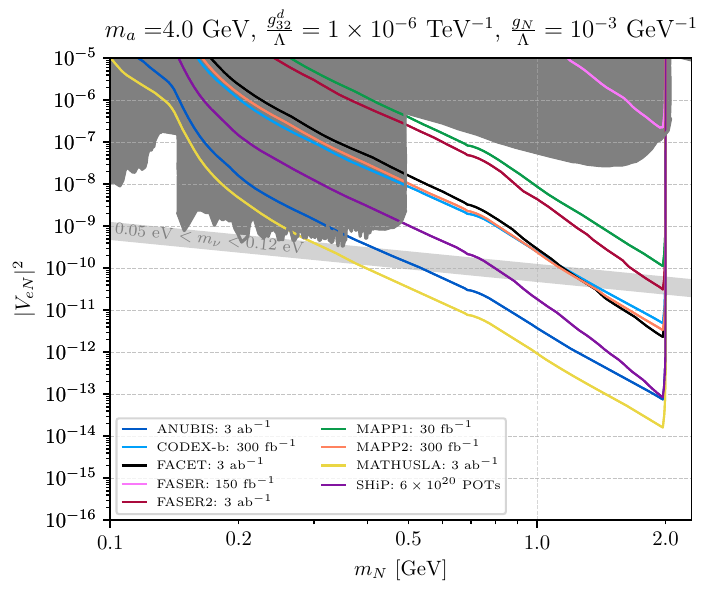}
	\caption{Sensitivity reach of the LHC far detectors and SHiP for the $B$-scenario, shown in the plane $|V_{eN}|^2$ vs.~$m_N$, for $m_a=1.0, 2.5,$ and 4.0 GeV.
    We have fixed $\frac{g^d_{3,2}}{\Lambda}=4\times 10^{-6}$ TeV$^{-1}$ $(1\times 10^{-6}$ TeV$^{-1}$) for $m_a=1.0$ and 2.5 GeV (4.0 GeV) and $\frac{g_N}{\Lambda} = 10^{-3}$ GeV$^{-1}$.
    In the plot of $m_a=1.0$ GeV, the sensitivity reach of FASER lies above the shown upper limit of the $y$-axis.
 }
 \label{fig:sensitivity_B_1em3}
\end{figure}

The numerical results for $g_N/\Lambda=10^{-3}$ GeV$^{-1}$ are presented in Fig.~\ref{fig:sensitivity_D_1em3} and Fig.~\ref{fig:sensitivity_B_1em3} for the $D$- and $B$-scenarios, respectively.
All the plots are shown in the $(m_N, |V_{eN}|^2)$ plane, for different choices of $m_a$, while the coupling $\frac{g^u_{2,1}}{\Lambda}$ is fixed at $2\times 10^{-4}$ TeV$^{-1}$, and $\frac{g^d_{3,2}}{\Lambda}=4\times 10^{-6}$ TeV$^{-1}$ or $1\times 10^{-6}$ TeV$^{-1}$ depending on the ALP mass.
We display the currently excluded parameter regions in the dark gray area, considering the existing leading bounds on the minimal-scenario HNL~\cite{PIENU:2017wbj,Bryman:2019bjg,CHARM:1985nku,NA62:2020mcv,T2K:2019jwa,Barouki:2022bkt}.
Further, we show a light-gray band which corresponds to the model parameter space that could explain the non-vanishing active-neutrino mass with the type-I seesaw relation $|V_{eN}|^2\simeq m_{\nu}/m_N$, for the active-neutrino mass $m_{\nu}$ between 0.05 eV and 0.12 eV.
The upper and lower bounds on the active-neutrino mass are respectively determined from neutrino-oscillation experiments~\cite{Canetti:2010aw} and cosmological observations~\cite{Planck:2018vyg}.

In addition, as noted in Sec.~\ref{sec:intro}, some existing bounds on the HNLs in the minimal scenario can be reinterpreted in terms of the theoretical scenarios we study here.
In particular, for the present study, such bounds include only those from CHARM~\cite{Boiarska:2021yho} and BEBC~\cite{Barouki:2022bkt}, for the $D$-scenario.\footnote{For the $B$-scenario, only sub-leading bounds obtained at Belle~\cite{Belle:2013ytx} considered $B$-meson decays and thus should, in principle, be reinterpreted, too. However, the Belle search requires a prompt lepton, which is absent in our $B$-scenario, and the search is hence not considered for reinterpretation.}
For this reason, we specifically show these bounds as red and cyan solid curves, respectively, in Fig.~\ref{fig:sensitivity_D_1em3}, and one observes that they are the leading ones at $m_N \gtrsim 0.5$ GeV.
Given that the two searches provide similar bounds on $|V_{eN}|^2$ in the kinematically relevant range of $m_N$, we choose to reinterpret the bounds from CHARM in terms of our model, following the simple recast method discussed in Ref.~\cite{Beltran:2023nli}.
The result is displayed as a red dashed curve in each plot of Fig.~\ref{fig:sensitivity_D_1em3}.
As expected, the reinterpreted bounds are largely strengthened compared to the original ones derived for the minimal scenario only.
They show that the CHARM search~\cite{CHARM:1985nku} has excluded large parameter regions.

In all these presented plots, we observe that some of the considered experiments can probe certain regions in the parameter space not only beyond the current bounds (by up to 10 orders of magnitude in some cases), but also covering large portions of the type-I-seesaw band.
For the $D$-scenario, SHiP is estimated to be sensitive to the lowest values of $|V_{eN}|^2$, leading MATHUSLA and ANUBIS.
The superiority of SHiP is primarily due to its largest production rates of the charm mesons and its high acceptance to LLPs originating from charm-meson decays as a result of its forward position with respect to the beam dump.
The excellent performance expected at MATHUSLA and ANUBIS, results, respectively, from a gigantic volume and the proximity from the IP. For the $B$-scenario, the strongest experiments are MATHUSLA and ANUBIS, followed by SHiP.
Now the SHiP experiment is estimated to be less sensitive than MATHUSLA and ANUBIS, as a consequence of the relatively small production rates of the $B$-mesons at its beam dump.

The upper mass reach of the HNL is $m_a/2$ in all cases, and therefore, a heavier ALP implies enhanced upper reach to $m_N$.
If an experiment is sensitive to heavier HNLs, it can exclude smaller values of the mixing angle, essentially as a result of the decoupling of the production and decay of the HNL in our theoretical model.
Consequently, we find that in each scenario, the plot for the heaviest ALP shows the best sensitivity reach to $|V_{eN}|^2$ attained at $m_N\lesssim m_a/2$.
Specifically, in the $D$-scenario with $m_a=1.5$ GeV, the SHiP experiment can exclude values of $|V_{eN}|^2$ down to as low as $10^{-17}$ with the sensitivity of MATHUSLA being weaker by $\lesssim 100$.
In the $B$-scenario with $m_a=2.5\text{ and }4.0$ GeV, the MATHUSLA, ANUBIS, and SHiP experiments can be sensitive to $|V_{eN}|^2\sim \mathcal{O}(10^{-14})$.
We note that for smaller values of the ALP QFV coupling with the quarks, the sensitivity reaches will be weakened.
Concretely, the signal-event number is proportional to $\Big(\frac{g^u_{2,1}}{\Lambda}\Big)^2$ or $\Big(\frac{g^d_{3,2}}{\Lambda}\Big)^2$, and to $|V_{eN}|^2$ in the large decay-length limit (essentially the small $|V_{eN}|^2$ regime).
Therefore, if, for example, the ALP QFV couplings are reduced by a factor of 10, the sensitivity reach to the lowest values of $|V_{eN}|^2$ should be impaired by factor of about 100.

\begin{figure}[t]
	\centering
	\includegraphics[width=0.495\textwidth]{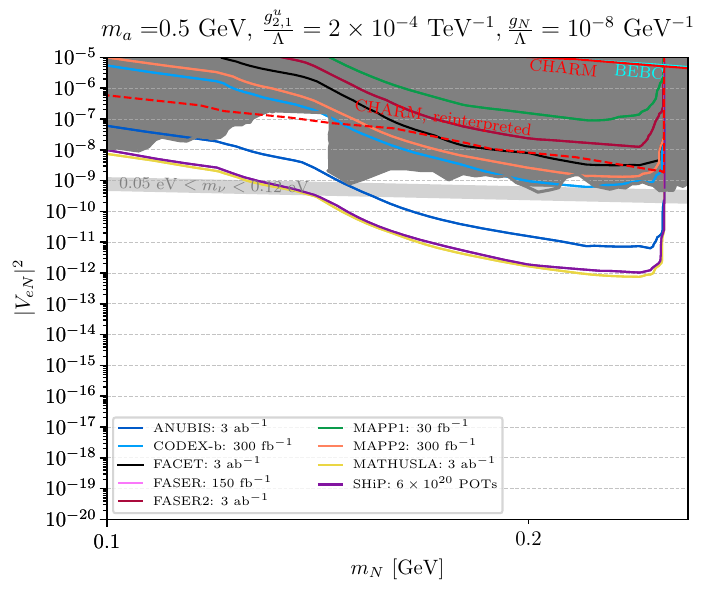}
	\includegraphics[width=0.495\textwidth]{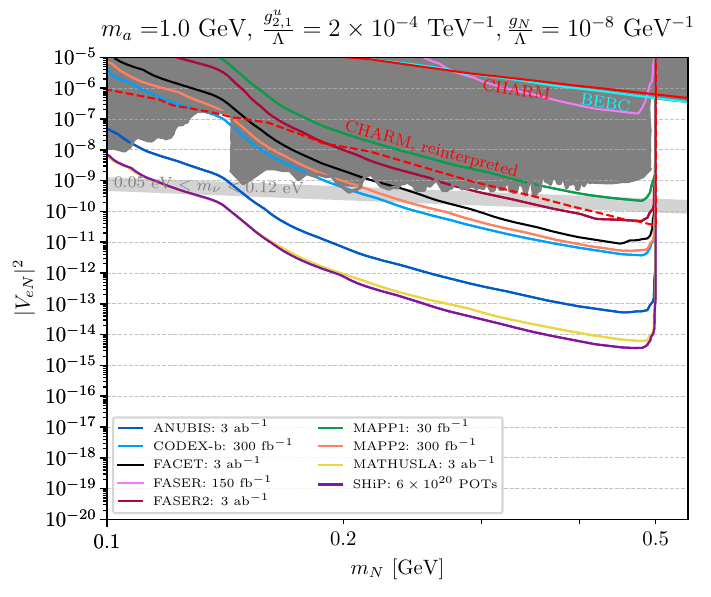}
	\includegraphics[width=0.495\textwidth]{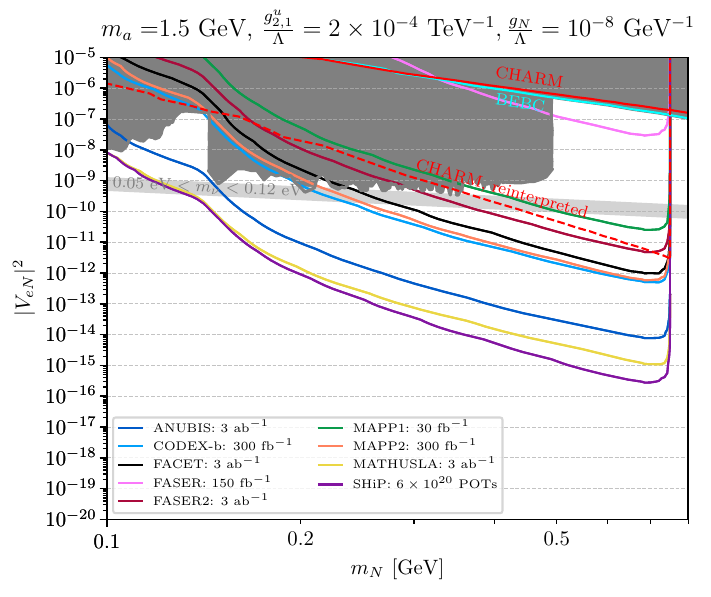}
	\caption{The same as Fig.~\ref{fig:sensitivity_D_1em3} but for $g_N/\Lambda=10^{-8}$ GeV$^{-1}$.
                Now the ALP is long-lived at all the considered LHC far-detector and the SHiP experiments.
                The displayed reinterpreted CHARM bounds are derived under the assumption of a promptly decaying ALP; nevertheless, we choose to present them for reference purpose.
 }
 \label{fig:sensitivity_D_1em8}
\end{figure}

\begin{figure}[t]
	\centering
	\includegraphics[width=0.495\textwidth]{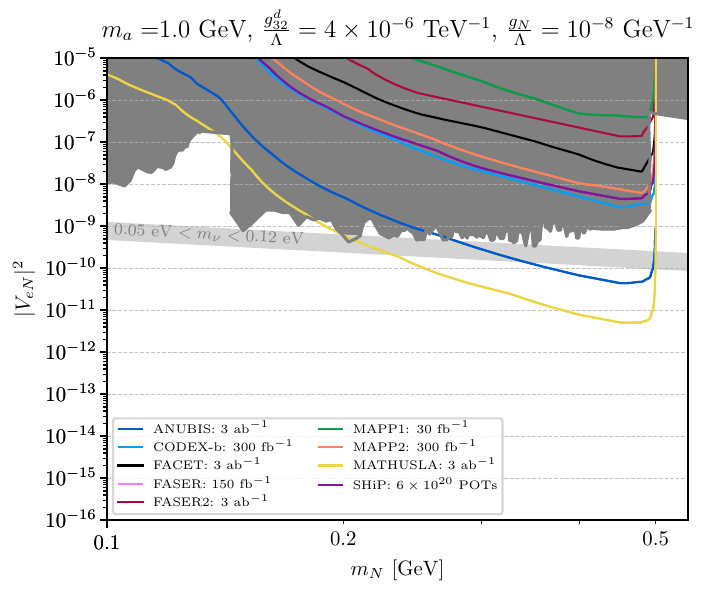}
	\includegraphics[width=0.495\textwidth]{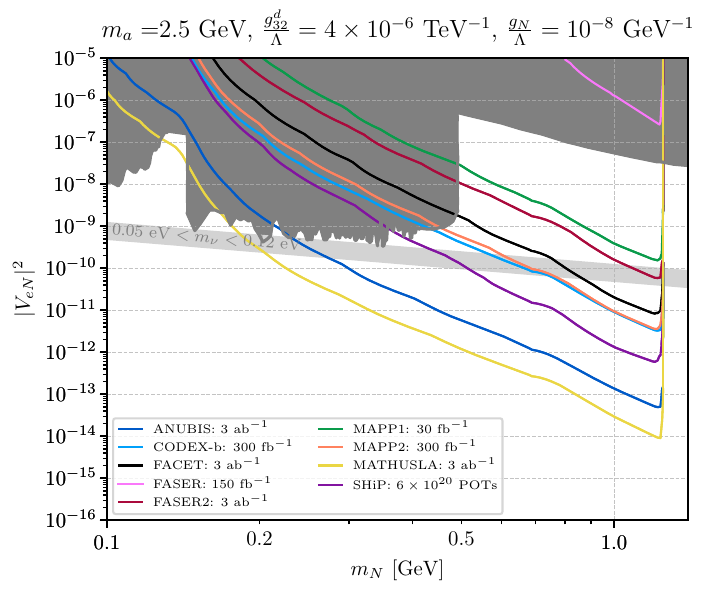}
	\includegraphics[width=0.495\textwidth]{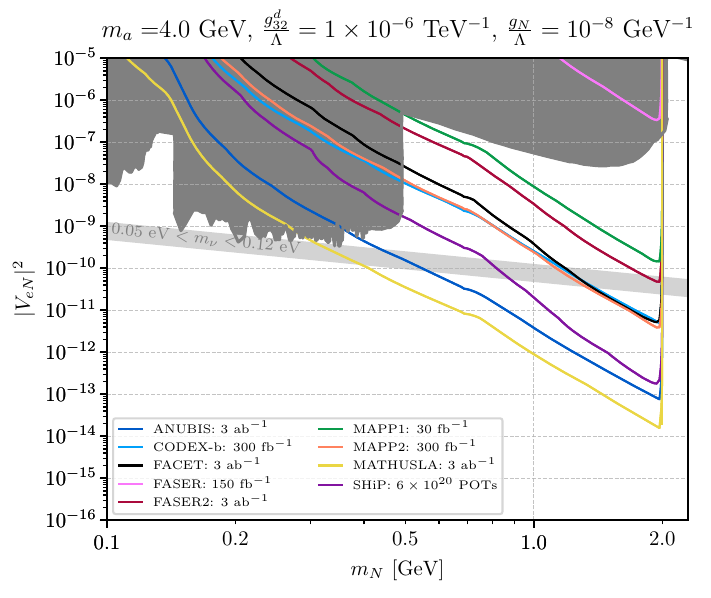}
	\caption{The same as Fig.~\ref{fig:sensitivity_B_1em3} but for $g_N/\Lambda=10^{-8}$ GeV$^{-1}$.
 }
 \label{fig:sensitivity_B_1em8}
\end{figure}

In Fig.~\ref{fig:sensitivity_D_1em3} and Fig.~\ref{fig:sensitivity_B_1em3}, we have assumed $g_N/\Lambda = 10^{-3}$ GeV$^{-1}$.
In fact, for $g_N/\Lambda$ between $10^{-6}$ GeV$^{-1}$ and $10^{-1}$ GeV$^{-1}$, as we numerically find out, the ALP is always promptly decaying at both the LHC and the SHiP experiments, and the corresponding sensitivity results are essentially identical.
For even smaller values of $g_N/\Lambda$, however, the ALP becomes long-lived, varying the expected sensitivity reach.
To illustrate this effect, in Fig.~\ref{fig:sensitivity_D_1em8} and Fig.~\ref{fig:sensitivity_B_1em8}, we show the numerical results for $g_N/\Lambda=10^{-8}$ GeV$^{-1}$.
We observe that compared to the case of $g_N/\Lambda=10^{-3}$ GeV$^{-1}$, now the sensitivity reach of all the considered experiments to the lowest values of $|V_{eN}|^2$ has reduced to different extents.
Concretely, the reduction effect is more prominent for the benchmarks of a lighter ALP, as the result of a longer lifetime.
Nevertheless, orders of magnitude beyond the present bounds can be probed by many of these experiments.

We note that in Fig.~\ref{fig:sensitivity_D_1em8}, we show the existing bounds from CHARM and BEBC as we do in Fig.~\ref{fig:sensitivity_D_1em3}, since such bounds should be reinterpreted.
However, given the long lifetime of the ALP here, it is not simple to reinterpret these bounds, unless we perform a full recast and reinterpretation of the CHARM and BEBC experiments.
Nonetheless, we decide to present the reinterpreted CHARM bounds assuming a short-lived ALP here, for reference purpose.

We conclude that these results show the huge potential of the LHC far detectors and the SHiP experiment in searching for long-lived HNLs produced in decays of the ALPs from pseudoscalar mesons.

\section{Conclusions}\label{sec:conclusions}

In this work, we have considered a model of the ALP coupled to quarks off-diagonally and to a pair of HNLs.
Both the ALP and the HNLs are well-motivated candidates as the mediator particle between the SM and a hidden sector of DM.
For simplicity and for the purpose of phenomenological studies, we have assumed in this work that only one HNL is kinematically relevant and it mixes with the electron neutrino only\footnote{For the HNL mixed with the muon neutrino, the sensitivity results should be similar, except for the kinematical thresholds' effects arising form the mass difference between the electron and the muon.  However, if the HNL mixes with the tau neutrino, the sensitivity results should be weakened to a larger extent, as a result of the much larger mass of the $\tau$-lepton.}.

We choose two benchmark scenarios with the QFV coupling $g^u_{2,1}$ and $g^d_{3,2}$, leading, respectively, to bottom and charm pseudoscalar mesons' decays into an ALP associated with a lighter (pseudoscalar or vector) meson.
Here, the two QFV couplings are fixed at values allowed by leading experimental measurements performed at tau-charm and $B$ factories.
We then assume the ALP coupling with the HNLs at representative values of $g_N/\Lambda = 10^{-3}$ GeV$^{-1}$ or $g_N/\Lambda = 10^{-8}$ GeV$^{-1}$, both obeying the perturbativity requirement.
We find that with these two values of $g_N/\Lambda$ the ALP decays promptly or after traveling a macroscopic distance, respectively, into a pair of the HNLs, for the kinematically relevant ranges of $m_a$ and $m_N$.
Further, for this range of $m_N$, the HNLs are necessarily long-lived, considering the present bounds on their mixing parameters with the active neutrinos.
We thus have investigated the sensitivity reach of the LHC far detectors and the recently approved SHiP experiment to the long-lived HNLs produced in the above-mentioned signal process.
We have performed MC simulations with state-of-the-art tools in order to determine the acceptance of the considered experiments to these long-lived HNLs, and then computed the expected rates of the signal events of reconstructed DVs at each experiment.
Since these experiments are (mostly) estimated to suffer from vanishing background events for such DV searches, we have shown 3-signal-event isocurves in the plane $|V_{eN}|^2$ vs.~$m_N$ for various choices of $m_a$ in each scenario, as the exclusion bounds at 95\% C.L.

In the numerical results, we present the sensitivity reach of the LHC far detectors and SHiP.
Compared with the existing bounds on $|V_{eN}|^2$ with the HNLs in the minimal scenario, we observe that orders of magnitude beyond these bounds can be probed, especially by SHiP, MATHUSLA, and ANUBIS.
In the $D$-scenario, we reinterpret the CHARM bounds on an HNL in the minimal scenario~\cite{Boiarska:2021yho}, in terms of our ALP-HNL model, and find that the CHARM search~\cite{CHARM:1985nku} already excludes large parts of the parameter space, but is still orders of magnitude weaker than the sensitivity reach of several future experiments we study here.
Further, in these sensitivity plots, we display a band corresponding to the region targeted by the type-I seesaw mechanism for explaining the light-neutrino mass between 0.05 eV and 0.12 eV with the HNL.
We observe that large parts of this band can be covered by most of the included experiments for both $D$- and $B$-scenarios and both considered values of $g_N/\Lambda$, thus providing strong motivation for performing the DV searches at these experiments.
It is worth noting that for the case of $g_N/\Lambda = 10^{-8}$ GeV$^{-1}$ with which the ALP is long-lived at all the considered experiments, the sensitivity reach is generally weakened to some extent.

We briefly discuss the case with presence of flavor-diagonal couplings of the ALP with the SM quarks.
If additionally we assume non-vanishing flavor-diagonal couplings similarly-sized to the flavor-violating couplings we have taken, the flavor-diagonal couplings can both loop-induce contributions to the tree-level flavor-violating ones and modify the ALP's decay widths.
The former contributions are negligible compared to the tree-level flavor-violating couplings themselves given their similar strengths, but the latter contributions may be important.
We numerically check and find that in the ``prompt'' case with $g_N/\Lambda = 10^{-3}$ GeV$^{-1}$ the decay width $\Gamma(a\to N N)$ stays orders of magnitude larger than that of hadronic decays of the ALP mediated via the diagonal couplings across the range of the kinematically allowed ALP mass.
However, in the ``long-lived'' case with $g_N/\Lambda = 10^{-8}$ GeV$^{-1}$ the hadronic decay width of the ALP becomes comparable to or dominant over $\Gamma(a\to N N)$ depending on the masses of the ALP and the HNL, and as a result the ALP may remain long-lived or turn promptly decaying and its decay branching ratio into a pair of the HNLs can be reduced to distinct extents.

We also comment that in models with both a light scalar boson that mixes with the SM-like Higgs boson and a dark photon which can, e.g.~acquire mass via a Higgs mechanism with the light scalar boson, a similar signal-process topology exists.
Here, the light scalar particle can be produced in e.g.~$B\to K$ transitions and subsequently decay to a pair of long-lived dark photons.
We refer to Refs.~\cite{Araki:2020wkq,Li:2021rzt,Foguel:2022unm,Araki:2022xqp,Cheung:2024oxh,Araki:2024uad} for some recent studies in this scenario.

Before closing, we discuss the outlook for some follow-up studies of this work, listed below:
\begin{enumerate}
    \item We have focused on charm and bottom mesons' decays in this paper.
    In principle, the ALP can also stem from kaon decays.
    For the corresponding mass range of the HNL, the DUNE experiment to be launched at Fermilab~\cite{DUNE:2020lwj,DUNE:2020ypp,DUNE:2020jqi,DUNE:2021cuw,DUNE:2021mtg} is expected to provide the strongest constraining power among the present and upcoming experiments; see e.g.~Refs.~\cite{Ballett:2019bgd,Gunther:2023vmz,Beltran:2023ksw}.
    Such a search for the long-lived HNLs performed at DUNE should cover the mixing parameters multiple orders of magnitude beyond the type-I seesaw band, for the theoretical scenario of the ALP coupled to the strange and down quarks and to a pair of the HNLs.

    \item We have considered only the case where the ALP is produced on shell.
    Including the off-shell ALP contributions could enhance the sensitivities~\cite{Araki:2024uad}.

    \item In addition to the LHC far detectors and SHiP, the $B$-meson scenario can be studied for $B$ factories such as Belle II~\cite{Belle-II:2010dht,Belle-II:2018jsg}.
    Since it is an ongoing experiment, a detailed study including detector efficiencies can be performed, leading to more realistic results.
    Similarly, the $D$-scenario can be probed at tau-charm factories such as BESIII~\cite{BESIII:2009fln,Huang:2022wuo,BESIII:2020nme} and STCF~\cite{Achasov:2023gey}.

    \item Besides the $P\to P'/V a, a\to N N$ signal process which we have investigated, the considered QFV couplings of the ALP could induce neutral-meson decays into a pair of $N$'s via an $s$-channel ALP: $P\to a \to N N$, with $P$ being the $D^0$ or $B_s^0$ meson.

\end{enumerate}

\section*{Acknowledgment}

We thank Arsenii Titov for useful discussions.
Y.Z.~is supported by the National Natural Science Foundation of China under Grant No.~12475106 and the Fundamental Research Funds for the Central Universities under Grant No.~JZ2023HGTB0222.
W.L.~is supported by National Natural Science Foundation of China (Grant No.~12205153).

\bibliographystyle{JHEP}
\bibliography{bib}

\end{document}